# Structural and Superconducting Properties in the Te-doped Spinel $CuRh_2Se_4$

*Kuan Li[a], Lingyong Zeng[a], Longfu Li[a], Rui Chen[a], Peifeng Yu[a], Kangwang Wang[a], Chao Zhang[a], Zaichen Xiang[a], Huixia Luo[a,b,c,d*]*

[a] School of Materials Science and Engineering, Sun Yat-sen University, No. 135, Xingang Xi Road, Guangzhou, 510275, P. R. China

[b] State Key Laboratory of Optoelectronic Materials and Technologies, Sun Yat-sen University, No. 135, Xingang Xi Road, Guangzhou, 510275, P. R. China

[c] Key Lab of Polymer Composite & Functional Materials, Sun Yat-sen University, No. 135, Xingang Xi Road, Guangzhou, 510275, P. R. China

[d] Guangdong Provincial Key Laboratory of Magnetoelectric Physics and Devices, Sun Yat-sen University, No. 135, Xingang Xi Road, Guangzhou, 510275, P. R. China

**ABSTRACT** In this paper, we discuss the impact of tellurium (Te) doping on the spinel superconductor $CuRh_2Se_4$. We conducted a comprehensive evaluation of the structural and superconducting properties of the system using various techniques, including X-ray diffraction (XRD), resistivity, magnetization, and specific heat measurements. Based on our XRD analysis, we found that the spinel superconductor $CuRh_2Se_{4-x}Te_x$ ($0 \leq x \leq 0.28$) crystallizes in the space group $Fd\bar{3}m$ (227), while the layered compound $CuRh_2Se_{4-x}Te_x$ ($2.8 \leq x \leq 4.0$) crystallizes in the space group $P\bar{3}m1$ (164). The upper critical magnetic field can be increased from 0.95(2) T for $CuRh_2Se_4$ to 3.44(1) T for $CuRh_2Se_{3.72}Te_{0.28}$ by doping with elemental Te. However, the layered compound $CuRh_2Se_{4-x}Te_x$ ($2.8 \leq x \leq 4.0$) did not exhibit superconducting properties. Besides, the specific heat measurements of $CuRh_2Se_{4-x}Te_x$ ($x = 0, 0.1, 0.28$) indicate that the Te element doping affects the electronic structure and interactions of the material and breaks the stability of the superconducting pairing, which leads to a decrease in the $T_c$. Finally, we show the electronic phase diagram of $T_c$ with Te doping to summarise our findings.

**INTRODUCTION**

Spinel is a type of material that has great potential for various applications. It has excellent electrical, magnetic, and thermal properties, making it relevant in many fields. Spinel exhibits unique physical properties, including multiferroicity, magnetostriction, and Kerr rotation [1-6], which have attracted much attention from researchers. The spinel structure is described by $AB_2X_4$, where metal ions occupy the A and B positions, and the X position is occupied by oxygen, sulfur, selenium, or tellurium. Spinel compounds have the potential for various technological applications due to their fascinating physical characteristics. Despite discovering several thousand spinels, only a tiny fraction of them have demonstrated superconductivity.

Spinel is a crucial superconductor material that holds great potential for applications in various fields [7-29]. The cubic spinel compounds $CuRh_2Se_4$ and $CuRh_2S_4$ were reported to exhibit superconductivity in 1967 by van Maaren et al. [7] and Robbins et al. [8], with superconducting transition temperature ($T_c$) of 3.4 K and 4.7 K, respectively. After a few years, the bulk superconducting oxide spinel $LiTi_2O_4$ was discovered and had a $T_c$ of 13.7 K, which set a record for the highest $T_c$ among the spinel compound family [9]. Recently, superconductivity has also been observed in $MgTi_2O_4$ oxide spinel films grown on $MgAl_2O_4$ substrates, achieved through superlattice engineering of $MgTi_2O_4$ and $SrTiO_3$ [10]. Other spinel superconductors that have been reported, including electron-doped $CuIr_2S_4$ and $CuIr_2Se_4$, as well as the undoped ternary sulpospinel superconductors, $CuCo_2S_4$ [11-19, 27, 28]. The $CuV_2S_4$ compound, which has a cubic spinel structure composed of copper and sulfur, exhibits three distinct charge density wave (CDW) states at different temperatures: $T_{CDW1}$ at 50 K, $T_{CDW2}$ at 75 K, and $T_{CDW3}$ at 90 K. These different states suggest that the phase transition in this compound may be caused by the behavior of electrons within its structure. And the $T_c$ of $CuV_2S_4$ is 3.20 - 4.45 K [21, 22]. Recently, our group found that Pt doping can significantly enhance the upper critical magnetic field, in which the upper critical magnetic field $\mu_0H_{c2}(0)$ increased from 0.6 T for the undoped $CuRh_2Se_4$ sample to 4.93 T for the Pt-doped composition $Cu(Rh_{0.94}Pt_{0.06})Se_4$ [29].

On the other hand, the introduction of Tellurium (Te) significantly influences both the crystal structure and the properties of the superconducting materials. For example, Te doping significantly impacts the critical temperature $T_c$, critical current density $J_c$, and flux pinning energy $U$ of FeSe single crystals. Suitable Te doping can increase the $J_c$ and $U$. Nevertheless, excessive Te doping leads to an increase in non-superconducting areas, which results in a swift reduction in both flux pinning energy and critical current density [30]. The $T_c$ of $Fe_ySe_{1-x}Te_x$ films is intricately linked to the chemical composition, with Fe vacancies playing a crucial role in promoting higher $T_c$ values [31]. The magnetic field dependence of the resistive transition in polycrystalline $\alpha$-$FeSe_{1-x}$ and $FeSe_{1-x}Te_x$ suggests that Te doping causes the system to become more two-dimensional and significantly enhances the upper critical field value, whereas the FeSe system behaves more like a 3D superconductor where Te replaces Se enlarging the lattice, increasing the $T_c$, and also greatly increasing the $\mu_0H_{c2}(0)$. This great enhancement may stem from the more 2D-like properties in the tellurium-doped system.

When α-FeSe is doped with 70 % Te, $\mu_0H_{c2}(0)$ is enhanced to 96.9 Tesla, well beyond the Pauli paramagnetic limit, estimated to be $\mu_0H^{\text{Pauli}} = 4kT/\pi\mu_0$ = 28.79 Tesla, providing more insight into the unconventional pairing mechanism that plays a role in superconductors [32]. In the previous study, Te-doped 2H-NbSe$_2$ can cause a new type of superconducting material, 1T-NbTeSe [33], or new compositions 2H-NbSe$_{2-x}$Te$_x$ [34]. These findings suggest that Te doping can significantly change the structural or physical properties.

Motivated by the research above, we propose to use Te doping spinel CuRh$_2$Se$_4$ and thus, design a series of CuRh$_2$Se$_{4-x}$Te$_x$. In this paper, we investigated the structure and physical properties of CuRh$_2$Se$_{4-x}$Te$_x$ by X-ray diffraction, resistivity, magnetic susceptibility, and specific heat measurements.

**EXPERIMENT**

Polycrystalline CuRh$_2$Se$_{4-x}$Te$_x$ samples were fabricated through a conventional solid-state reaction method. Stoichiometric ratios of Cu (99.9%), Rh (99.95%), Se (>99.9%), and Te (99.999%) were combined in quartz tubes. Subsequently, the quartz tubes underwent a ten-day heating process at 850 ℃ within a high vacuum setting (< 1×10$^{-1}$ MPa). The resulting powder samples were meticulously reprocessed, pelletized, and sintered at 850 ℃ for 48 hours.

The crystal structure was confirmed using XRD analysis. Data was collected from 10º to 100º with a step width of 0.01 º and a constant scan speed of 1 º/min at room temperature, utilizing the MiniFlex instrument from Rigaku with Cu Kα radiation. The XRD patterns were refined via the Rietveld method using the FULLPROF suite package software. To characterize the morphology and microstructure, scanning electron microscopy (SEM) and energy-dispersive X-ray spectroscopy (EDS) analysis were conducted using the EVO system from Zeiss. For further analysis, resistivity measurements, zero-field-cooling (ZFC) magnetic susceptibility studies, and specific heat evaluations were carried out using the Quantum Design physical property measurement system DynaCool. Heat capacity measurements in the range of 1.8 K - 6 K and electrical resistivity assessments on rectangular samples (4.5 × 1.3 × 0.6 mm³) were performed using the four-probe method on the Quantum Design physical property measuring system (PPMS). Additionally, magnetic susceptibilities were measured using the PPMS. The $T_c$ was determined from the specific heat capacity $C_p(T)$ through the equal entropy construction.

**RESULTS AND ANALYSIS**

The series of polycrystalline XRD spectra and correlation analysis of CuRh$_2$Se$_{4-x}$Te$_x$ are shown in **Fig. 1**. XRD analysis at room temperature indicates that all CuRh$_2$Se$_{4-x}$Te$_x$ ($0 \leq x \leq 0.28$) samples are cubic phase, and the space group is $Fd\bar{3}m$ (227). Room temperature XRD Rietveld refinement of all samples revealed a strong correlation with the spinel CuRh$_2$Se$_4$ crystal structure (PDF card number: 04-0058682), although minor RhSe$_2$ impurities were detected in some samples. On the other hand, XRD results show that the powder diffraction pattern of the CuRh$_2$Se$_{4-x}$Te$_x$ ($0 \leq x \leq 0.28$) sample is mainly spinel phase with space group $Fd\bar{3}m$ (227). The trigonal phase begins to appear in the 0.28 < x < 2.8 region, indicating that the spinel phase and the layered trigonal phase coexist. The main layered trigonal phase with space group $P\bar{3}m1$ (164) is obtained in the doped region of $2.8 \leq x \leq 4.0$. The peaks around 35 ° were zoomed as depicted on the right side of **Fig. 1(a)**. When the doping amount of Te increases gradually, the obvious peaks shift to the left. According to Bragg's law $2dSin\theta = n\lambda$, it is not difficult to explain that because the ionic radius of Te is larger than that of Se, the crystal plane spacing gradually increases with the increase of Te doping amount. **Fig. 1(b)** illustrates the lattice parameters exhibiting a remarkable linear correlation with Te doping content. **Fig. 1(c)**, the XRD Rietveld refinement of the sample $CuRh_2Se_{3.85}Te_{0.15}$ at room temperature is presented. Subsequently, employing Thompson-Cox-Hastings pseudo-Voigt peak shapes, we conduct a quantitative analysis of the XRD data using the FULLPROF software to determine the lattice parameters. The lattice parameter increased from 10.2645(4) Å for $CuRh_2Se_4$ to 10.3092(5) Å for $CuRh_2Se_{3.72}Te_{0.28}$. Further refined data for other compositions in the system can be found in **Fig. S2**. **Fig. 1(c)** internal shows the crystal structure of $CuRh_2Se_{4-x}Te_x$ ($0 \leq x \leq 0.28$), where the Te element partially replaces the Se element. Furthermore, the atomic ratios obtained from EDXS confirmed that the synthesized compounds were very close to the target compositions, as shown in **Table S1**. Additionally, EDXS mapping, as shown in **Fig. S1**, indicated that the polycrystalline samples exhibited uniformity. The XRD pattern of the highly doped level ($2.8 \leq x \leq 4.0$) of the layered compound $CuRh_2Se_{4-x}Te_x$ is shown in **Fig. 2 (a)**. It can be observed that the peak around 31 ° is amplified. As the doping level of Te gradually increases, a noticeable shift of the peak to the left can be observed. Similarly, according to Bragg's law $2dSin\theta = n\lambda$, it can be easily explained that as the doping level of Te increases, the interplanar spacing gradually increases due to the smaller ionic radius of Se compared to Te. All CuRh$_2$Se$_{4-x}$Te$_x$ (2.8 ≤ $x$ ≤ 4.0) samples crystallized in the space group $P\bar{3}m1$ (164). The relationship between the lattice parameter and Te doping content is well represented by a linear function, as depicted in **Fig. 2(b)**. At room temperature, a representative sample of CuRh$_2$Se$_{1.2}$Te$_{2.8}$ was analyzed via XRD Rietveld refinement, as **Fig. 2(c)** displays the results. Then, the XRD data were quantitatively analyzed using FULLPROF software to determine the lattice parameters. The lattice parameter increased from 3.8307(3) Å for CuRh$_2$Se$_{1.2}$Te$_{2.8}$ to 4.0116(2) Å for CuRh$_2$Te$_4$. In **Fig. 2(c)**, the internal view depicts the crystal structure of CuRh$_2$Se$_{4-x}$Te$_x$, where the value of $x$ ranges from 2.8 to 4.0. Additionally, Cu ions are inserted between layers.

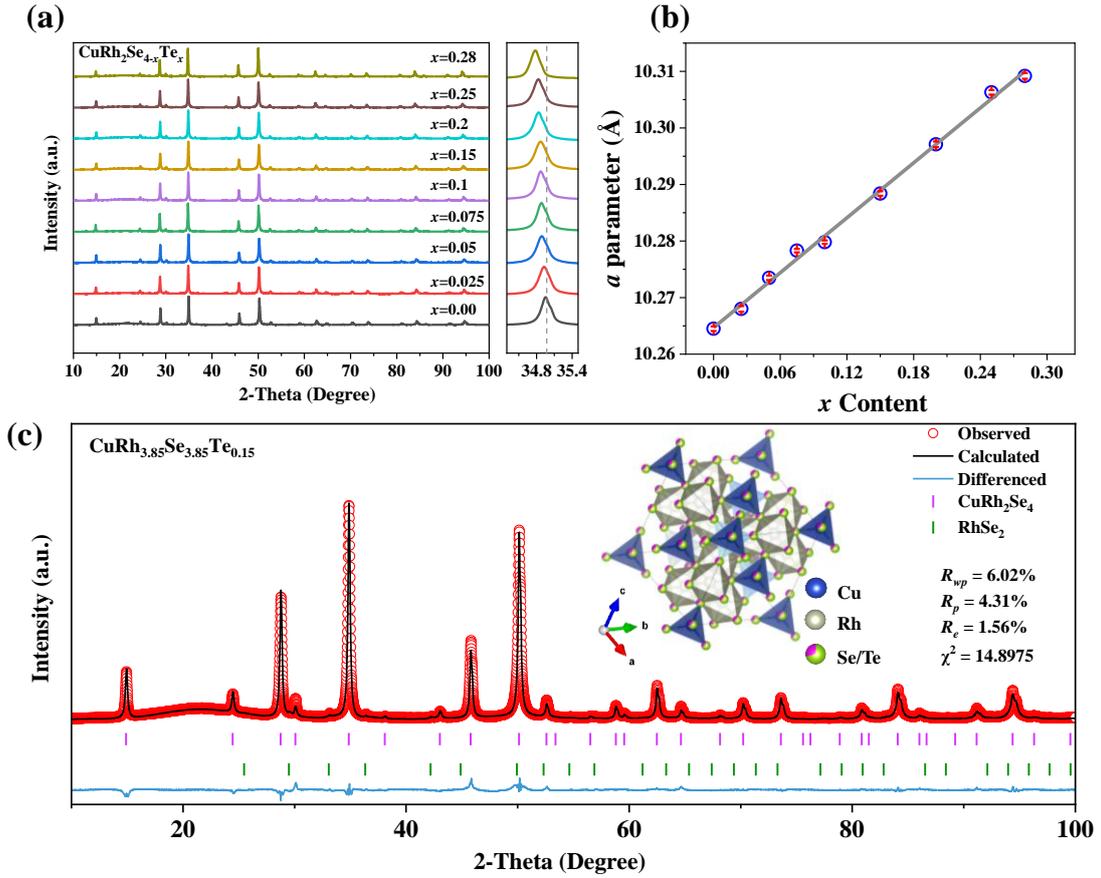

**Fig. 1.** Structural characterization and analysis of the spinel $CuRh_2Se_{4-x}Te_x$ ($0 \leq x \leq 0.28$). (a) Powder X-ray diffraction patterns for the spinel $CuRh_2Se_{4-x}Te_x$ ($0 \leq x \leq 0.28$). (b) Variation of the calculated lattice parameters $a$ with Te-doping (c) Rietveld refinement profile of the XRD of the spinel $CuRh_2Se_{3.85}Te_{0.15}$, the crystal structure for the spinel $CuRh_2Se_{4-x}Te_x$ ($0 \leq x \leq 0.28$) samples with space group $Fd\bar{3}m$ is shown in Fig. 1c.

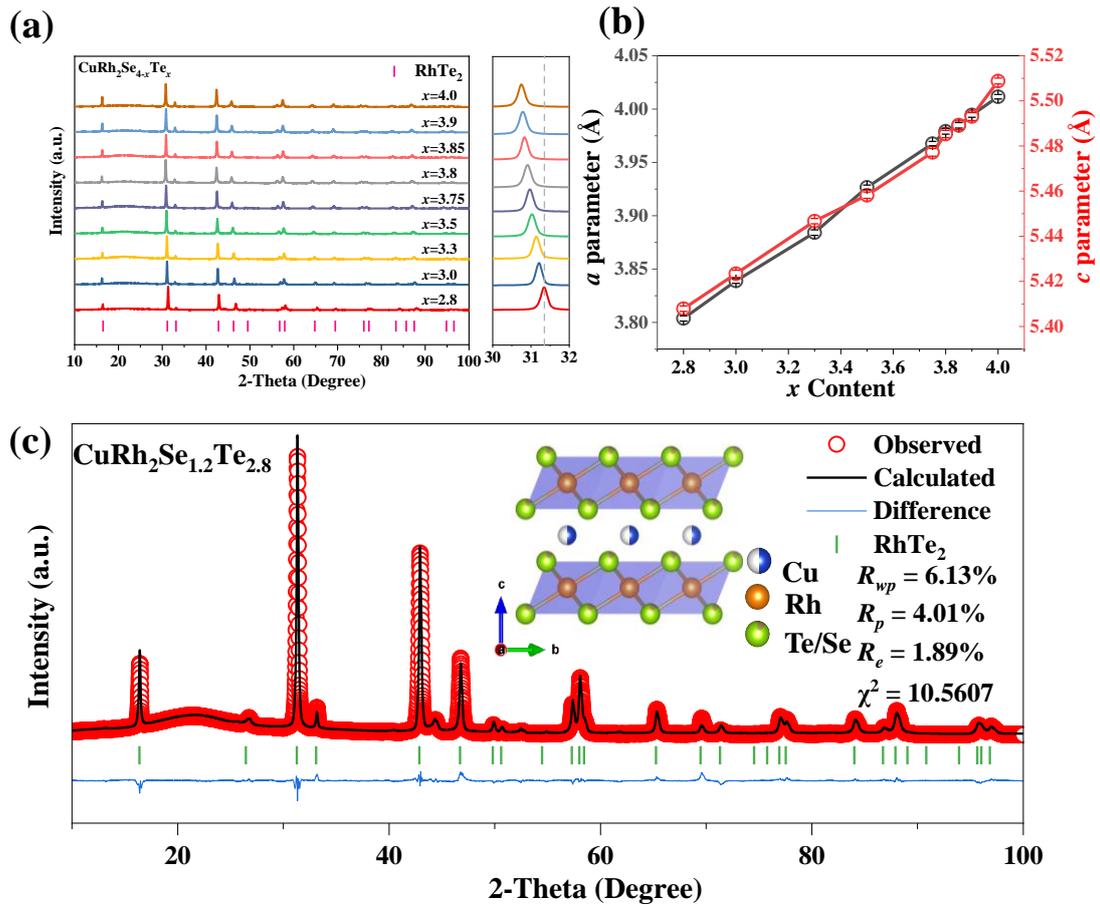

**Fig. 2.** Structural characterization and analysis of the layer compound CuRh$_2$Se$_{4-x}$Te$_x$ (2.8 ≤ x ≤ 4.0). (a) Powder XRD patterns for the layer compound CuRh$_2$Se$_{4-x}$Te$_x$ (2.8 ≤ x ≤ 4.0). (b) Variation of the calculated lattice parameters *a* and c with Te-doping. (c) Rietveld refinement profile of the XRD of the layer compound CuRh$_2$Se$_{1.2}$Te$_{2.8}$, the crystal structure for the layer compound CuRh$_2$Se$_{4-x}$Te$_x$ (2.8 ≤ x ≤ 4.0) samples with space group $P\bar{3}m1$ is shown in Fig. 2c.

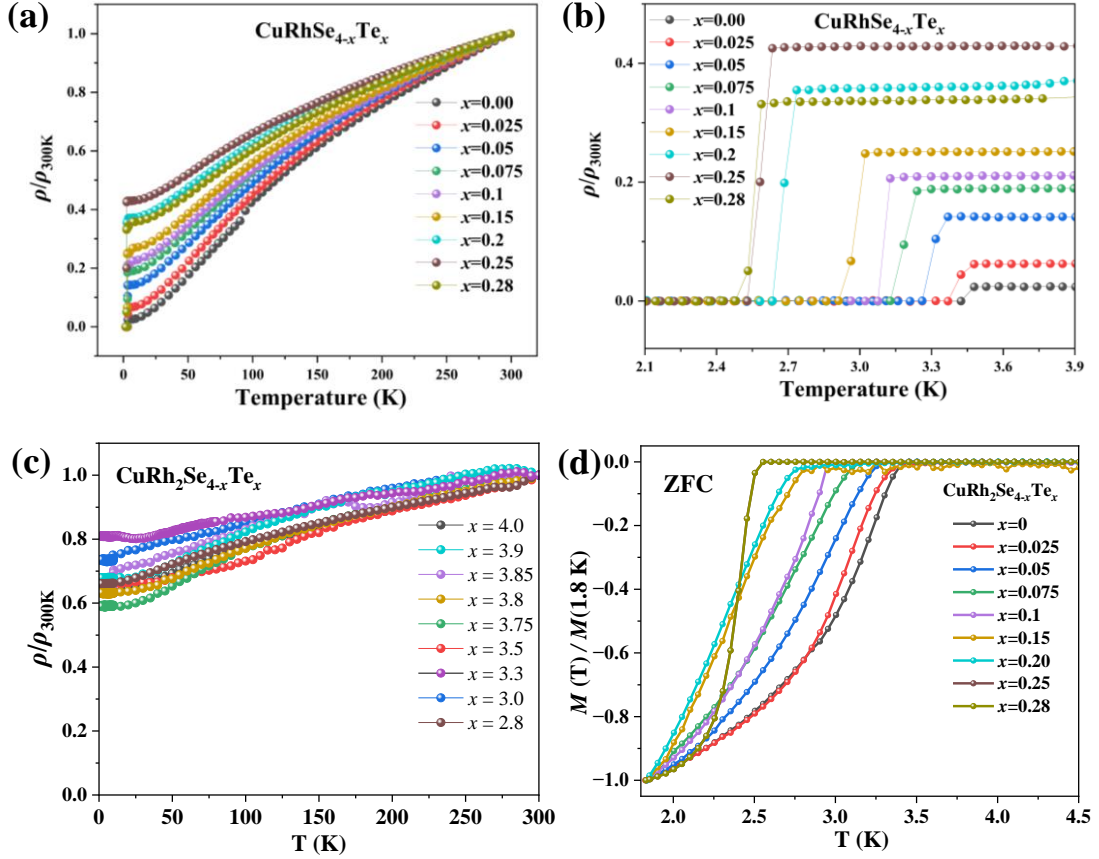

**Fig. 3.** (a) Temperature dependence of normalized $\rho/\rho_{300K}$ of CuRh$_2$Se$_{4-x}$Te$_x$ ($0 \leq x \leq 0.28$). (b) Temperature dependence of normalized $\rho/\rho_{300K}$ of CuRh$_2$Se$_{4-x}$Te$_x$ ($0 \leq x \leq 0.28$) at 2.1 – 3.9 K. (c) Temperature dependence of normalized $\rho/\rho_{300K}$ of CuRh$_2$Se$_{4-x}$Te$_x$ ($2.8 \leq x \leq 4.0$). (d) Magnetic susceptibilities for the spinel CuRh$_2$Se$_{4-x}$Te$_x$ ($0 \leq x \leq 0.28$) samples at the superconducting transitions under a 30 Oe magnetic field.

The resistivity of CuRh$_2$Se$_{4-x}$Te$_x$ polycrystalline samples was measured at temperatures ranging from 300 K to 1.8 K. For the CuRh$_2$Se$_{4-x}$Te$_x$ sample, the highest $T_c$ of 3.5 K was observed in the parent CuRh$_2$Se$_4$, which is also consistent with that reported in the previous literature [13, 29]. However, superconductivity is inhibited after Te doping and regularly decreases with increasing Te content. **Fig. 3(a)** provides a detailed illustration of the superconducting transition across a wide temperature range from 1.8 K to 300 K, whereas **Fig. 3(b)** depicts the resistivity trend at lower temperatures spanning from 2.1 to 3.9 K. The $T_c$ was defined as a resistivity of 50 %. When $x = 0$, $T_c$ reaches a maximum of 3.5 K; when $x = 0.15$, $T_c$ gradually decreases to about 3 K with the increase of Te content, and when $x$ reaches the maximum doping level of 0.28, $T_c$ finally drops to about 2.5 K. **Fig. S3** shows the phase plot of the change in superconducting transition temperature $T_c$ with the amount of Te doping. *RRR* is generally defined as the ratio of resistance measured at 273 K (freezing point of water) and 4.2 K (boiling point of helium at standard atmospheric pressure) [35]. Here, we define the residual resistivity ratio at 300 K and 5 K (*RRR* = $R_{300K}/R_{5K}$). **Table S2** summarizes the $T_c$ and *RRR* values of $CuRh_2Se_{4-x}Te_x$ ($0 \leq x \leq 0.28$) samples. **Fig. S4** displays the *RRR* of each $CuRh_2Se_{4-x}Te_x$ sample. The parent sample showed a rapid transition to a superconducting state at 3.5 K, with an initial *R*-value of 41, indicating a high degree of homogeneity in the undoped sample. The *RRR* values indicate that Te doping introduces disorder in the $CuRh_2Se_{4-x}Te_x$ system, suppressing the critical temperature $T_c$. In addition, the subsequent decrease in *RRR* also indicates an enhanced effect of electron scattering. In the temperature range of 300 K to 1.8 K, the resistivity of polycrystalline samples of $CuRh_2Se_{4-x}Te_x$ ($2.8 \leq x \leq 4.0$) was measured. However, no superconducting transition was observed in the samples when the temperature was lowered to 1.8 K, as shown in **Fig. 3(c)**. The ZFC measurements under 30 Oe magnetic field were applied to detect the diamagnetism of superconducting $CuRh_2Se_{4-x}Te_x$ ($0 \leq x$

≤ 0.28) samples in the temperature range of 1.8 - 4.5 K as depicted in **Fig. 3(d)**. The decreasing trend of spinel CuRh$_2$Se$_{4-x}$Te$_x$ (0 ≤ $x$ ≤ 0.28) compound. As a result, $T_c$ decreases almost to 3.05 K at $x$ = 0.1. Significantly, $T_c$ also reduces with increasing doping content of Te. The trend of $T_c$ shows great agreement with the R-T results at slightly lower levels due to the suppression of the applied magnetic field. However, the $T_c$ is not as steep as expected. We suggest that this phenomenon originates from the decay of the Meissner screen current, which is common in polycrystalline samples. Next, we systematically studied the spinel CuRh$_2$Se$_{4-x}$Te$_x$ (0 ≤ $x$ ≤ 0.28). We also performed upper and lower critical magnetic field fitting tests on polycrystalline samples.

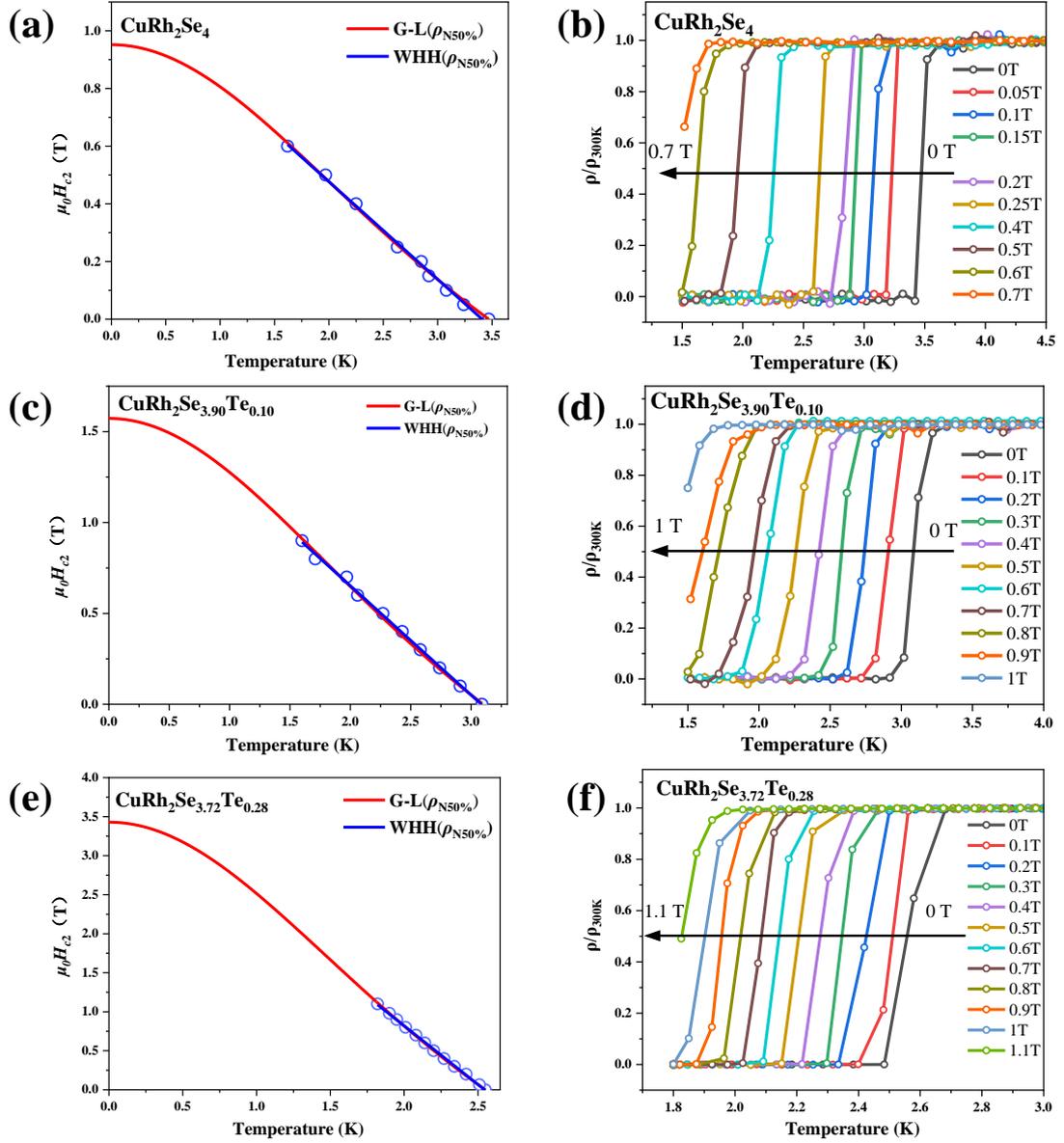

Fig. 4. The temperature dependence measurement of the upper critical $\mu_0H_{c2}(0)$ fields for CuRh$_2$Se$_{4-x}$Te$_x$ ($x$ = 0, 0.1, 0.28); (a), (c), and (e) show the refinements of CuRh$_2$Se$_{4-x}$Te$_x$ ($x$ = 0, 0.1, 0.28), respectively. The red curve exhibits the refinement by G-L theory, while the blue curve displays the refinement by the WHH model; (b), (d), and (f) show the detailed process for determining $\mu_0H_{c2}(0)$.

The upper critical field $\mu_0H_{c2}(0)$ was systematically investigated by R-T measurements. The measurement procedure for CuRh$_2$Se$_{4-x}$Te$_x$ ($x$ = 0, 0.1, 0.28) is shown in **Fig. 4(a)-4(f)**. The distribution of points exhibits good function compliance, as demonstrated in **Fig. 4(a)**, **4(c)**, and **4(e)**. Werthamer-Helfand-Hohenburg (WHH) and Ginzberg-Landau (G-L) theories of $\mu_0H_{c2}(0)$ theory are based on data plots for the $\mu_0H_{c2}(0)$, which are established using the 50% criterion of the normal state resistivity value. From **Fig. 4(b)**, **4(d)**, and **4(f)**, it can be seen that $T_c$ exhibits a significant shift toward the low-temperature region as the applied magnetic field strength increases. The $d\mu_0H_{c2}(0)/dT$ is the slope of the plot around $T_c$ used in the simplified WHH equation: $\mu_0H_{c2}(0) = -0.693T_c(\frac{d\mu_0H_{c2}}{dT})|_{T = T_c}$. As dirty limits for the WHH model, the calculated values for CuRh$_2$Se$_4$, CuRh$_2$Se$_{3.9}$Te$_{0.1}$, and CuRh$_2$Se$_{3.72}$Te$_{0.28}$ are 0.89(2) T, 1.29(3) T, and 2.60(2) T, respectively. Due to the Bowling Limit effect, the $\mu_0H_{c2}(0)$ calculated by the WHH model has to be smaller than the Pauling limit field $\mu_0H^{Pauli} = 1.86T_c$ [36]. In this case, the $\mu_0H^{Pauli}$ is 6.32(2) T, 5.76(2) T, and 4.65(3) T for CuRh$_2$Se$_4$, CuRh$_2$Se$_{3.9}$Te$_{0.1}$, and CuRh$_2$Se$_{3.72}$Te$_{0.28}$, respectively. The $\mu_0H_{c2}(0)$ also follows a functional distribution based on G-L theory: $\mu_0H_{c2}(T) = \mu_0H_{c2}(0) \times \frac{(1-(T/T_c)^2)}{(1+(T/T_c)^2)}$, where $\mu_0H_{c2}(0)$ can be calculated. The estimated values of $\mu_0H_{c2}(0)$ for CuRh$_2$Se$_4$, CuRh$_2$Se$_{3.9}$Te$_{0.1}$, and CuRh$_2$Se$_{3.72}$Te$_{0.28}$ are 0.95(2) T, 1.57(4) T, and 3.44(1) T, respectively. The results indicate that replacing Se with Te can greatly raise the $\mu_0H_{c2}(0)$, even though there is a minor difference between the values of $\mu_0H_{c2}(0)$ derived using WHH modeling and G-L theory. We think that Te substitution effects are the source of the rise in the $\mu_0H_{c2}(0)$. The *RR*R depicted in **Fig.**

**S4** serves as a measure of disorder, where a diminishing value signifies heightened disorder. Upon doping the Te element into the pristine $CuRh_2Se_4$ spinel, a stark decrease in the *RRR* value is evident, illustrating the effective role of Te as a scattering center. Thus, the inclusion of Te elements notably exacerbates the disorder. This leads to intensified electron scattering and a reduction in the mean free path of carriers [37-40]. Based on these factors, we hypothesize that the shorter coherence length resulting from impurity scattering may be the reason for the augmentation of $\mu_0 H_{c2}(0)$ in Te-doped samples. Then, we measured the M-H curves at various temperatures to find the lower critical transition magnetic field $\mu_0 H_{c1}(0)$. The parent sample $CuRh_2Se_4$ with the greatest $T_c$ was chosen for testing. The details of the $\mu_0 H_{c1}(0)$ test is displayed in **Fig. S5**. All values with corresponding temperatures are shown in **Fig. S5(a)** and correspond to the equation $\mu_0 H_{c1}^*(T) = \mu_0 H_{c1}^*(0)(1-(T/T_c)^2)$, which is indicated by the red solid line. The calculated $\mu_0 H_{c1}(0)$ is 143(6) Oe, which is close to the value previously reported in the literature (100 Oe) [29]. In addition, the lower critical field is calculated by the equation $\mu_0 H_{c1}(0) = \mu_0 H_{c1}*(0)/(1-N)$ when considering an *N* value of 0.5. Thus, the modified lower critical field $\mu_0 H_{c1}(0)$ is 286 Oe. The inset of **Fig. S5(a)** shows the field-dependent magnetization intensity measurements of the spinel $CuRh_2Se_4$ superconductor at different temperatures below $T_c$. The magnetization intensity *M* and the external magnetic field *H* show a linear association in the range of low external magnetic field strengths, as indicated by the equation $M_{fit} = e + fH$. In this case, *f* stands for the linear relationship's slope, and *e* for the intercept [40]. **Fig. S5(b)** shows the *M*- $M_{fit}$ curve. The field where the magnetization intensity starts to deviate from the linear response is the uncorrected lower critical field $\mu_0 H_{c1}^*(0)$ for that temperature.

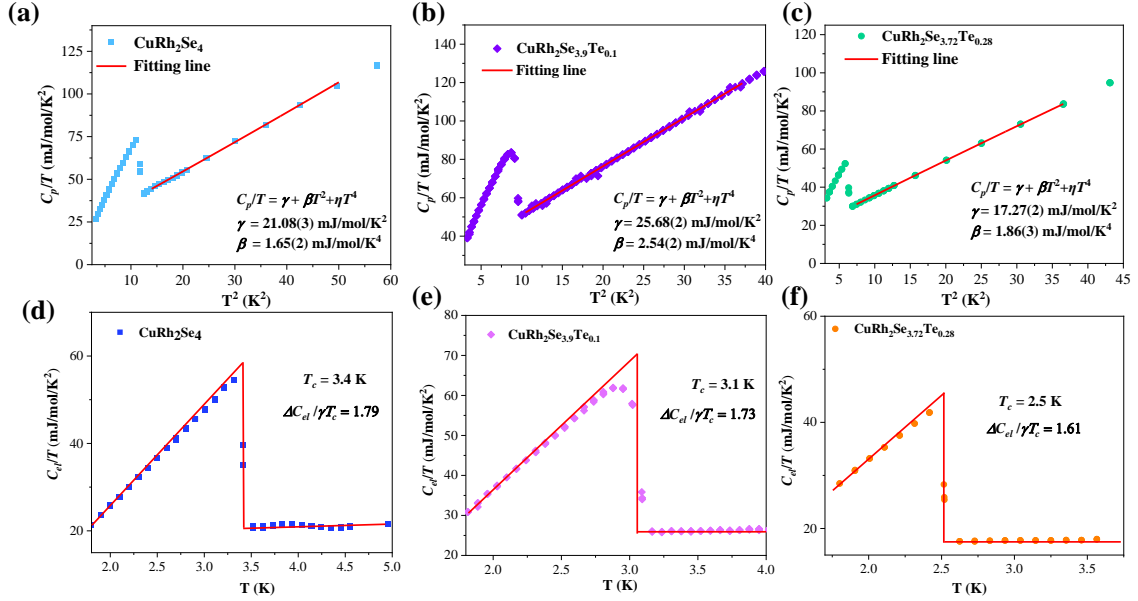

**Fig. 5** The temperature-dependent specific heat for the spinel $CuRh_2Se_{4-x}Te_x$ ($x$ = 0, 0.1, 0.28). (a), (b), and (c) The $C_p/T$ vs $T^2$ curves, fitted with the equation $C_p/T = \gamma + \beta T^2 + \eta T^4$. (d), (e), and (f) Electronic contribution to the heat capacity divided by temperature ($C_{el}/T$) vs temperature $T$ without an applied magnetic field.

After observing the zero resistivity and the Meissner effect during magnetization, we further performed specific heat measurements of $CuRh_2Se_{4-x}Te_x$ ($x$ = 0, 0.1, 0.28), as shown in **Fig. 5**. The heat capacity data above $T_c$ can be fitted by the equation $C_p/T = \gamma + \beta T^2 + \eta T^4$, where two-term $\beta T^2 + \eta T^4$ are used to express the phonon contribution, and $\gamma$ represents the normal state electronic specific heat coefficient. The values of $\gamma$ and $\beta$ for the $CuRh_2Se_4$ are 21.08(3) mJ mol$^{-1}$ K$^{-2}$ and 1.65(2) mJ mol$^{-1}$ K$^{-4}$, respectively. In addition, the values of $\gamma$ and $\beta$ for the $CuRh_2Se_{3.72}Te_{0.28}$ are 17.27(2) mJ mol$^{-1}$ K$^{-2}$ and 1.86(3) mJ mol$^{-1}$ K$^{-4}$, respectively. The **Figs. 5(d), (e), and (f)** display the $C_{el}/T$ vs $T$ curves. The estimated $T_c$ obtained from the equal-entropy construction agrees with the resistivity and magnetization measurements of the polycrystalline sample. The normalized specific heat jump value $\Delta C_{el}/\gamma T_c$ is 1.79 for CuRh$_2$Se$_4$ and 1.61 for CuRh$_2$Se$_{3.72}$Te$_{0.28}$, which is slightly above the Bardeen - Cooper - Schrieffer (BCS) weak coupling limit value of 1.43, indicating the presence of bulk superconductivity. The Debye temperature ($\Theta_D$) is calculated as $\Theta_D = (12\pi^4 nR/5\beta)^{1/3}$, where $n$ is the number of atoms in the unit of the formula, and $R$ is the molar gas constant. The calculated $\Theta_D$ is 202 K for CuRh$_2$Se$_4$ and 194 K for CuRh$_2$Se$_{3.72}$Te$_{0.28}$. Given the $\Theta_D$ and $T_c$, the electron-phonon coupling constant $\lambda_{ep}$ of CuRh$_2$Se$_4$ and CuRh$_2$Se$_{3.72}$Te$_{0.28}$ can be calculated to be 0.64 and 0.59 with $\mu* = 0.13$ using McMillan's equation:

$$\lambda_{ep} = \frac{1.04 + \mu^* \ln\left(\frac{\Theta_D}{1.45 T_c}\right)}{(1 - 1.62\mu^*) \ln\left(\frac{\Theta_D}{1.45 T_c}\right) - 1.04}$$, respectively. [41]. Finally, the density of states (DOSs) located at the Fermi energy level $N(E_F)$ can also be estimated using the formula $N(E_F) = \frac{3}{\pi^2 k_B^2 (1 + \lambda_{ep})}\gamma$ with values $\gamma$ and $\lambda_{ep}$. Since the parent sample CuRh$_2$Se$_4$ is close to CuRh$_2$Se$_{3.9}$Te$_{0.1}$ in terms of $T_c$ and $N(E_F)$, we chose to compare the parent sample CuRh$_2$Se$_4$ with the highest doped sample CuRh$_2$Se$_{3.72}$Te$_{0.28}$. The $N(E_F)$ of 5.5 states/eV f.u. for the parent CuRh$_2$Se$_4$ is significantly higher than that of 4.6 states/eV f.u. for the highest doped CuRh$_2$Se$_{3.72}$Te$_{0.28}$. This suggests that the doping of the Te element may change the electronic band structure of the material, including tuning the parameters of the DOSs near the Fermi energy level and electronic bandwidth. Such changes may affect the superconducting pairing mechanism and hence the $T_c$. In addition, **Table 1** summarizes the parameters of the superconducting properties of CuRh$_2$Se$_{4-x}$Te$_x$ ($x$=0, 0.1, 0.28) and the spinel compounds CuRh$_2$S$_4$, CuRh$_2$Se$_4$ and CuRh$_{1.88}$Pt$_{0.12}$Se$_4$ doped samples from other papers. Furthermore, the Te elemental doping in the polycrystalline

samples of $CuRh_2(S_{1-x}Te_x)_4$ ($0 \leq x \leq 0.1$) leads to a decrease in the $T_c$. Once Te is introduced, the lattice constant $a$ increases. There is a strong correlation between the $T_c$ and the lattice constant $a$. The suppression of $T_c$ is most likely due to the substitution of Te, which reduces the density of states at the $N(E_F)$ [42]. Besides, transport and magnetization strength measurements show that the $T_c$ of FeSe single crystals decreases with increasing Te doping. The overdoping of Te produces more non-superconducting regions, which rapidly reduces the flux pinning energy and critical current density [43]. Taken together, Te element doping may affect the electronic structure and interactions of the material and break the stability of the superconducting pairing, which leads to a decrease in the $T_c$.

**Table 1** summarizes the superconducting property parameters of spinel compounds $CuRh_2S_4$, $CuRh_2Se_4$, and $CuRh_2Se_4$ doped samples. The comparison of superconducting characteristic parameters for spinel compounds.

| Compound | $CuRh_2S_4$ [13] | $CuRh_{1.88}Pt_{0.12}Se_4$ [29] | $CuRh_2Se_4$ [this work] | $CuRh_2Se_{3.9}Te_{0.1}$ [this work] | $CuRh_2Se_{2.72}Te_{0.28}$ [this work] | $CuRh_2Se_4$ [29] |
|---|---|---|---|---|---|---|
| $T_c$ (K) | 3.5 | 3.84(2) | 3.4 | 3.11 | 2.5 | 3.38(1) |
| $\mu_0H_{c1}(0)$ (Oe) | - | 168(12) | - | 143(6) | - | 220(6) |
| $\mu_0H_{c2}(0)$ (T) ($\rho^{50\%}_N$ G-L theory) | - | 4.93(1) | 0.95(2) | 1.57(4) | 3.44(1) | 1.00(1) |
| $\mu_0H^{Pauli}$ | - | 7.176(4) | 6.32(2) | 5.76(2) | 4.65(3) | 6.311(2) |
| $-d\mu_0H_{c2}(0)/dT_c$ (T/K) | 0.614 | 1.418 | 0.347(3) | 0.601(5) | 1.503(2) | 0.352(8) |
| $\gamma$ (mJ mol$^{-1}$ K$^{-2}$) | 26.9 | - | 21.08(3) | 25.68(2) | 17.27(2) | 21.4 |
| $\beta$ (mJ mol$^{-1}$ K$^{-4}$) | - | - | 1.65(2) | 2.54(2) | 1.86(3) | - |
| $\Delta C_{el}/\gamma T_c$ | 1.89 | - | 1.79 | 1.73 | 1.61 | 1.68 |
| $\Theta_D$ (K) | 258 | - | 202 | 175 | 194 | 218 |
| $\lambda_{ep}$ | 0.66 | - | 0.64 | 0.65 | 0.59 | 0.63 |
| $N(E_F)$ (states/eV f.u.) | - | - | 5.5 | 6.6 | 4.6 | - |

The electronic phase diagrams of the spinel superconductor $CuRh_2Se_{4-x}Te_x$ ($0 \leq x \leq 0.28$) and high doping levels of layered compound $CuRh_2Se_{4-x}Te_x$ ($2.8 \leq x \leq 4.0$) are summarized in **Fig. 6**. The $T_c$ obtained from the R-T experimental results is summarized as a curve related to the Te doping concentration. As can be seen, $T_c$ shows a decreasing trend with increasing Te substitution, yet the $\mu_0 H_{c2}(0)$ enhances with the Te doping content in the spinel $CuRh_2Se_{4-x}Te_x$ ($0 \leq x \leq 0.28$) compositions. However, no superconductivity is found for the layered compound $CuRh_2Se_{4-x}Te_x$ ($2.8 \leq x \leq 4.0$) from 300 K to 1.8 K.

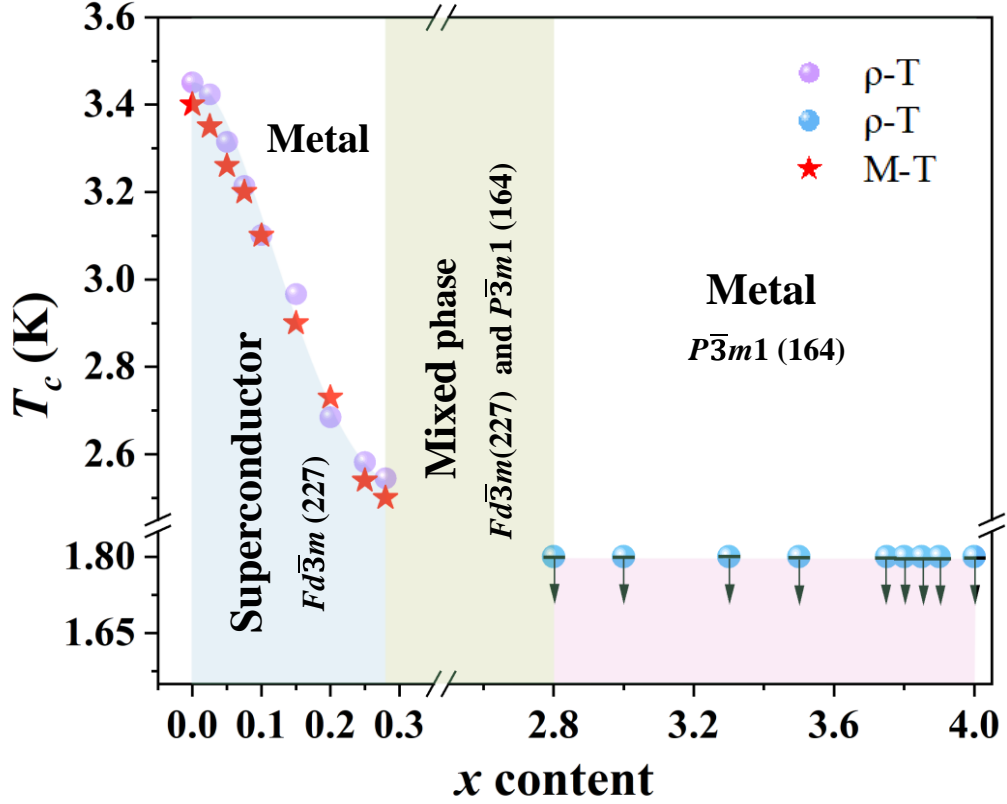

**Fig. 6** The electronic phase diagram for the spinel superconductor $CuRh_2Se_{4-x}Te_x$ ($0 \leq x \leq 0.28$) and the layered compound $CuRh_2Se_{4-x}Te_x$ ($2.8 \leq x \leq 4.0$) vs Te content.

## CONCLUSIONS

We have conducted a systematic study of the effect of Te doping on the spinel $CuRh_2Se_4$. XRD analysis revealed that the low-doping $CuRh_2Se_{4-x}Te_x$ ($0 \leq x \leq 0.28$) compositions crystallized in the spinel structure with the space group $Fd\bar{3}m$ (227), while the high-doping $CuRh_2Se_{4-x}Te_x$ ($2.8 \leq x \leq 4.0$) compounds crystallized in the layered structure with the space group $P\bar{3}m1$ (164). The XRD refinement data show that the lattice parameters increase with increasing Te doping. Resistivity and magnetization measurements show that $T_c$ decreases almost linearly with increasing Te content in the spinel $CuRh_2Se_{4-x}Te_x$ ($0 \leq x \leq 0.28$) system, whereas the highly doped levels of layered compound $CuRh_2Se_{4-x}Te_x$ (2.8 ≤ $x$ ≤ 4.0) did not exhibit superconducting properties. The *RRR* values indicate that Te doping introduces disorder in the $CuRh_2Se_{4-x}Te_x$ system. Notably, the highest-doped $CuRh_2Se_{3.72}Te_{0.28}$ spinel has an enhanced upper critical magnetic field $\mu_0H_{c2}(0)^{GL}$ = 3.44(1) T. Besides, the zero-field heat capacity indicates the presence of bulk superconductivity and the variation of the density of states is obtained. Our results reveal that Te doping can significantly modulate the superconductivity and structure of the spinel $CuRh_2Se_4$. Te elemental doping affects the electronic structure and interactions of the material and destabilizes the superconducting pairs, leading to a decrease in $T_c$. These findings can give new insight for further discovery of other new spinel superconductors or enhance the superconductivity of the original spinels.

**Notes**

The authors declare no competing financial interest

**ACKNOWLEDGMENT**

This work is supported by the National Natural Science Foundation of China (12274471, 11922415), Guangdong Basic and Applied Basic Research Foundation (2022A1515011168), 2024 Basic and Applied Basic Research Topic (Science and Technology Elite "Navigation" Project", 2024A04J6415). The experiments reported were conducted at the Guangdong Provincial Key Laboratory of Magnetoelectric Physics and Devices, No. 2022B1212010008.

REFERENCES

[1] J. Hemberger, H.A.K. von Nidda, V. Tsurkan, A. Loidl, Large Magnetostriction and Negative Thermal Expansion in the Frustrated Antiferromagnet $ZnCr_2Se_4$, Physical Review Letters 98(14) (2007) 147203.

[2] A.P. Ramirez, R.J. Cava, J. Krajewski, Colossal magnetoresistance in Cr-based chalcogenide spinels, Nature 386(6621) (1997) 156-159.
[3] Z. Yang, S. Tan, Z. Chen, Y. Zhang, Magnetic polaron conductivity in $FeCr_2S_4$ with the colossal magnetoresistance effect, Physical Review B 62(21) (2000) 13872-13875.
[4] S. Weber, P. Lunkenheimer, R. Fichtl, J. Hemberger, V. Tsurkan, A. Loidl, Colossal Magnetocapacitance and Colossal Magnetoresistance in $HgCr_2S_4$, Physical Review Letters 96(15) (2006) 157202.
[5] K. Ohgushi, T. Ogasawara, Y. Okimoto, S. Miyasaka, Y. Tokura, Gigantic Kerr rotation induced by a $d$-$d$ transition resonance in $MCr_2S_4$ ($M$ = Mn, Fe), Physical Review B 72(15) (2005) 155114.
[6] J. Hemberger, P. Lunkenheimer, R. Fichtl, H.A. Krug von Nidda, V. Tsurkan, A. Loidl, Relaxor ferroelectricity and colossal magnetocapacitive coupling in ferromagnetic $CdCr_2S_4$, Nature 434(7031) (2005) 364-367.
[7] N. H. Van Maaren, G.M. Schaeffer, F.K. Lotgering, Superconductivity in sulpho- and selenospinels, Phys. Lett. A 25(3) (1967) 238-239.
[8] M. Robbins, R.H. Willens, R.C. Miller, Superconductivity in the spinels CuRh2S4 and CuRh2Se4, Solid State Communications 5(12) (1967) 933-934.
[9] D.C. Johnston, H. Prakash, W.H. Zachariasen, R. Viswanathan, High temperature superconductivity in the Li Ti O ternary system, Mater. Res. Bull. 8(7) (1973) 777-784.
[10] W. Hu, Z. Feng, B.-C. Gong, G. He, D. Li, M. Qin, Y. Shi, Q. Li, Q. Zhang, J. Yuan, B. Zhu, K. Liu, T. Xiang, L. Gu, F. Zhou, X. Dong, Z. Zhao, K. Jin, Emergent superconductivity in single-crystalline $MgTi_2O_4$ films via structural engineering, Physical Review B 101(22) (2020) 220510.
[11] T. Bitoh, T. Hagino, Y. Seki, S. Chikazawa, S. Nagata, Superconductivity in Thiospinel $CuRh_2S_4$, Journal of the Physical Society of Japan 61(8) (1992) 3011-3012.
[12] M. Ito, J. Hori, H. Kurisaki, H. Okada, A.J.P. Kuroki, N. Ogita, M. Udagawa, H. Fujii, F. Nakamura, T. Fujita, T. Suzuki, Pressure-Induced Superconductor-Insulator Transition in the Spinel Compound $CuRh_2S_4$, Physical Review Letters 91(7) (2003)
[13] T. Shirane, T. Hagino, Y. Seki, T. Bitoh, S. Chikazawa, S. Nagata, Superconductivity in Selenospinel $CuRh_2Se_4$, Journal of the Physical Society of Japan 62(1) (1993) 374-375.
[14] R.N. Shelton, D.C. Johnston, H. Adrian, Measurement of the pressure dependence of $T_c$ for superconducting spinel compounds, Solid State Communications 20(11) (1976) 1077-1080.
[15] T. Hagino, Y. Seki, N. Wada, S. Tsuji, T. Shirane, K.-i. Kumagai, S. Nagata, Superconductivity in spinel-type compounds $CuRh_2S_4$ and $CuRh_2Se_4$, Physical Review B 51(18) (1995) 12673-12684.
[16] H. Suzuki, T. Furubayashi, G. Cao, H. Kitazawa, A. Kamimura, K. Hirata, T. Matsumoto, Metal-Insulator Transition and Superconductivity in Spinel-Type System $Cu_{1-x}Zn_xIr_2S_4$, Journal of the Physical Society of Japan 68(8) (1999) 2495-2497.
[17] G. Cao, H. Kitazawa, H. Suzuki, T. Furubayashi, K. Hirata, T. Matsumoto, Superconductivity in Zn-doped $CuIr_2S_4$, Physica C: Superconductivity 341-348 (2000) 735-736.

[18] H. Luo, T. Klimczuk, L. Müchler, L. Schoop, D. Hirai, M.K. Fuccillo, C. Felser, R.J. Cava, Superconductivity in the Cu(Ir$_{1-x}$Pt$_x$)$_2$Se$_4$ spinel, Physical Review B 87(21) (2013) 214510.
[19] Y.-Y. Jin, S.-H. Sun, Y.-W. Cui, Q.-Q. Zhu, L.-W. Ji, Z. Ren, G.-H. Cao, Bulk superconductivity and Pauli paramagnetism in nearly stoichiometric CuCo$_2$S$_4$, Physical Review Materials 5(7) (2021) 074804.
[20] T. Oda, M. Shirai, N. Suzuki, K. Motizuki, Electronic band structure of sulphide spinels CuM$_2$S$_4$ (M=Co, Rh,Ir), Journal of Physics: Condensed Matter 7(23) (1995) 4433.
[21] R.M. Fleming, F.J. DiSalvo, R.J. Cava, J.V. Waszczak, Observation of charge-density waves in the cubic spinel structure CuV$_2$S$_4$, Physical Review B 24(5) (1981) 2850-2853.
[22] N. Le Nagard, A. Katty, G. Collin, O. Gorochov, A. Willig, Elaboration, structure cristalline et proprietes physiques (transport, susceptibilité magnétique et R.M.N.) du spinelle CuV$_2$S$_4$, Journal of Solid State Chemistry 27(3) (1979) 267-277.
[23] T. Hagino, Y. Seki, S. Nagata, Metal - insulator transition in CuIr$_2$S$_4$ : Comparison with CuIr$_2$Se$_4$, Physica C: Superconductivity 235-240 (1994) 1303-1304.
[24] T. Furubayashi, T. Kosaka, J. Tang, T. Matsumoto, Y. Kato, S. Nagata, Pressure Induced Metal-Insulator Transition of Selenospinel CuIr$_2$Se$_4$, Journal of the Physical Society of Japan 66(5) (1997) 1563-1564.
[25] T. Furubayashi, T. Matsumoto, T. Hagino, S. Nagata, Structural and Magnetic Studies of Metal-Insulator Transition in Thiospinel CuIr$_2$S$_4$, Journal of the Physical Society of Japan 63(9) (1994) 3333-3339.
[26] T. Hagino, T. Tojo, T. Atake, S. Nagata, Metal-insulator transition at 230 K in a new thiospinel CuIr$_2$S$_4$, Philosophical Magazine B 71(5) (1995) 881-894.
[27] G. Oomi, T. Kagayama, I. Yoshida, T. Hagino, S. Nagata, Effect of pressure on the metal-insulator transition temperature in thiospinel CuIr$_2$S$_4$, Journal of Magnetism and Magnetic Materials 140-144 (1995) 157-158.
[28] G. Cao, T. Furubayashi, H. Suzuki, H. Kitazawa, T. Matsumoto, Y. Uwatoko, Suppression of metal-to-insulator transition and appearance of superconductivity in Cu$_{1-x}$Zn$_x$Ir$_2$S$_4$, Physical Review B 64(21) (2001) 214514.
[29] Y. He, Y.-X. You, L. Zeng, S. Guo, H. Zhou, K. Li, Y. Huang, P. Yu, C. Zhang, C. Cao, H. Luo, Superconductivity with the enhanced upper critical field in the Pt-doped CuRh$_2$Se$_4$ spinel, Physical Review B 105(5) (2022) 054513.
[30] R. Mao, Z. Wu, Z. Wang, Z. Pan, M. Xu, Z. Wang, Effect of Te-doping on the superconducting characteristics of FeSe single crystal, Journal of Alloys and Compounds 809 (2019) 151851.
[31] Y. Zhang, T. Wang, Z. Wang, Z. Xing, Effects of Te- and Fe-doping on the superconducting properties in Fe$_y$Se$_{1-x}$Te$_x$ thin films, Scientific Reports 12(1) (2022) 391.
[32] K. W. Yeh, H. C. Hsu, T. W. Huang, P. M. Wu, Y. L. Huang, T. K. Chen, J. Y. Luo, M. K. Wu, Se and Te Doping Study of the FeSe Superconductors, Journal of the Physical Society of Japan 77(Suppl.C) (2008) 19-22.

[33] D. Yan, S. Wang, Y. S. Lin, G. H. Wang, Y. J. Zeng, M. Boubeche, Y. He, J. Ma, Y. H. Wang, D. X. Yao, H. X. Luo, NbSeTe—a new layered transition metal dichalcogenide superconductor, *J. Phys.: Condens. Matter* 32 (2020) 025702.

[34] H.-T. Wang, L.-J. Li, D.-S. Ye, X.-H. Cheng, Z.-A. Xu, Effect of Te doping on superconductivity and charge-density wave in dichalcogenides 2H-NbSe$_{2-x}$Te$_x$ ($x$ = 0, 0.1, 0.2), Chinese Physics 16(8) (2007) 2471.

[35] J.D. Splett, D.F. Vecchia, L.F. Goodrich, A comparison of methods for computing the residual resistivity ratio of high-purity niobium, Journal of research of the National Institute of Standards and Technology 116(1) (2011) 489.

[36] A.M. Clogston, Upper Limit for the Critical Field in Hard Superconductors, Physical Review Letters 9(6) (1962) 266-267.

[37] X.-L. Wang, S.X. Dou, M.S.A. Hossain, Z.X. Cheng, X.Z. Liao, S.R. Ghorbani, Q.W. Yao, J.H. Kim, T. Silver, Enhancement of the in-field $J_c$ of MgB$_2$ via SiCl$_4$ doping, Physical Review B 81(22) (2010) 224514.

[38] K.S.B. De Silva, X. Xu, X.L. Wang, D. Wexler, D. Attard, F. Xiang, S.X. Dou, A significant improvement in the superconducting properties of MgB$_2$ by co-doping with graphene and nano-SiC, Scripta Materialia 67(10) (2012) 802-805.

[40] H. Luo, W. Xie, J. Tao, I. Pletikosic, T. Valla, G.S. Sahasrabudhe, G. Osterhoudt, E. Sutton, K.S. Burch, E.M. Seibel, J.W. Krizan, Y. Zhu, R.J. Cava, Differences in Chemical Doping Matter: Superconductivity in Ti$_{1-x}$Ta$_x$Se$_2$ but Not in Ti$_{1-x}$Nb$_x$Se$_2$, Chemistry of Materials 28(6) (2016) 1927-1935.

[41] W.L. McMillan, Transition Temperature of Strong-Coupled Superconductors, Physical Review 167(2) (1968) 331-344.

[42] N. Kijima, H. Yashiro, S. Nagata, Superconductivity in CuRh$_2$(S$_{1-x}$Te$_x$)$_4$ (0 ≤ $x$ ≤ 0.1), Journal of Physics and Chemistry of Solids 57(11) (1996) 1635-1639.

[43] R. Mao, Z. Wu, Z. Wang, Z. Pan, M. Xu, Z. Wang, Effect of Te-doping on the superconducting characteristics of FeSe single crystal, Journal of Alloys and Compounds 809 (2019) 151851.